\documentclass[aps,pra,showpacs,twocolumn]{revtex4}
\usepackage{amsmath, amssymb}
\usepackage{graphicx}
\usepackage{subfigure}

\begin{document}

\title{Entanglement signature in the mode structure of a single photon}

\author{C. Di Fidio}
\author{W. Vogel}
\affiliation{Arbeitsgruppe Quantenoptik, Institut f\"ur Physik, Universit\"at Rostock, D-18051 Rostock, Germany}

\date{\today}

\begin{abstract}

It is shown that entanglement, which is a quantum correlation property of at
least two subsystems, is imprinted in the mode structure of a single
photon. The photon, which is emitted by two coupled cavities, carries the
information on the concurrence of the two intracavity fields. This can be
useful for recording the entanglement dynamics of two cavity fields and for
entanglement transfer.

\end{abstract}


\maketitle

%%%%%%%%%%%%

An atom interacting with a quantized radiation-field mode 
in a high-$Q$ optical cavity plays an important role in quantum optics,
for a review see, e.g., Ref.~\cite{HarocheRaimond}.
The ability to coherently control individual quantum system,
and in particular the quantum control of single-photon 
emission from an atom in a cavity,
is a key requirement in various applications 
of quantum networks for distribution and processing of quantum
information~\cite{Cirac:3221,Brattke:3534,Knill:46,Monroe:238,Kimble:1023}.
Recently, single-photon sources operating
on the basis of adiabatic passage with just one atom
trapped in a high-$Q$ optical cavity have been 
realized~\cite{Parkins:3095,Hennrich:4872, Mckeever:1992, Hijlkema:253}.
In this way, the adjustment of the spatiotemporal profile of
single-photon pulses has been achieved
\cite{Kuhn:067901, Keller:1075}.
Moreover, the generation of single photons of
known circular polarization 
emitted into a well-defined spatiotemporal mode
has been possible \cite{Wilk:063601}, and
an atom-photon quantum interface 
involving atom-photon entanglement
has been realized~\cite{Wilk:488}.
More recently, the amplitude modulation
in the photon emission on a single 
atom-cavity system has been 
studied theoretically~\cite{Difidio:043822}
and experimentally~\cite{Bochmann:223601}.
In addition, photon-photon entanglement with a single trapped atom
in a high-finesse optical cavity has been performed~\cite{Weber:030501}.

In the present contribution,
in view of the widespread applications of
cavity-assisted single-photon sources, 
we study single-photon emission from 
a system consisting of two coupled atom-cavity subsystems in
a cascaded configuration~\cite{Carmichael:2273,Gardiner:2269}.
%We will show that 
The mode structure of the radiated
photon strongly depends on the entanglement 
%established 
between the two
intracavity fields and
%In addition, 
it sensitively depends on the presence or absence
of an atom in the second cavity.
We show how the entanglement of the intracavity fields can
be experimentally determined.

\begin{figure}
\includegraphics[width=7.0cm]{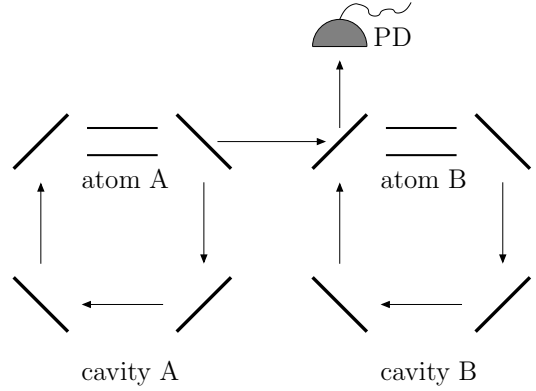}
\caption{The cascaded system
consists of two atom-cavity subsystems 
$A$ and $B$. A photodetector PD monitors the radiation 
field.}
\label{fig:figure_1}
\end{figure}

The system under study consists of two
atom-cavity subsystems $A$ and $B$,
where the source subsystem $A$ is cascaded with the
target subsystem $B$, cf. Fig.~\ref{fig:figure_1}.
The cavities have three perfectly reflecting mirrors and one
mirror with transmission coefficient $T \ll 1$.
In the two subsystems $A$ and $B$
we consider a two-level atomic transition
of frequency  $\omega_k$ (related to the atomic
energy eigenstates $|1_k\rangle$ and $|0_k\rangle$)
coupled to a cavity mode of frequency 
$\omega_k'$, where $k=a,b$ denotes the subsystem.
The cavity mode is detuned by $\Delta_k$
from the two-level atomic transition frequency, 
$\omega_k = \omega_k' + \Delta_k$, 
and  is damped by losses through the partially transmitting
cavity mirrors. 
In addition to the wanted outcoupling of the field,
the photon can be spontaneously emitted out the side
of the cavity into modes other than the one
which is preferentially coupled to the resonator.
Moreover, the photon may be absorbed or scattered
by the cavity mirrors.

To describe the dynamics of the system we 
use the following master equation for the reduced density 
operator $\hat \rho(t)$ of the system:
\begin{eqnarray}
\frac{d\hat\rho(t)}{dt} &=& \frac{1}{i\hbar} \! \left[\hat H,\hat\rho(t)\right] \!+\! \sum_{i = 1}^5
\left[ \hat J_i \hat\rho(t) \hat J_i^\dagger 
- \frac{1}{2} \hat J_i^\dagger \hat J_i \hat\rho(t) \right. \nonumber \\
&-& \left. \frac{1}{2} \hat\rho(t) \hat J_i^\dagger \hat J_i \right] .
\label{eq:master_1}
\end{eqnarray}
The Hamiltonian is given by
\begin{equation}
\hat H = \hat H_A + \hat H_B + i \hbar \frac{\sqrt{\kappa_a \kappa_b}}{2}
\left( e^{-i \phi} \hat b \hat a^\dagger - e^{i \phi} \hat b^\dagger \hat
a \right) \, ,
\label{eq:hamiltonian}
\end{equation}
where $\hat H_A$ and $\hat H_B$ describe the atom-cavity interaction in the two subsystems $A$ and $B$, respectively,
and, in the
rotating-wave approximation, are given by
\begin{equation}
\hat H_A = \hbar g_a  \left( \hat{a} \hat A_{10} 
+ \hat a^\dagger \hat A_{01}\right) + \hbar \Delta_a {\hat A}_{11}
   \, ,
\label{eq:JC_hamiltonian_A}
\end{equation}
and
\begin{equation}
\hat H_B = \hbar g_b \left( \hat{b} \hat B_{10} 
+ \hat b^\dagger \hat B_{01}\right) + \hbar \Delta_b {\hat B}_{11}
   \, .
\label{eq:JC_hamiltonian_B}
\end{equation}
The third term in Eq.~(\ref{eq:hamiltonian}) describes the
coupling between the two
cavities~\cite{Carmichael:2273,Gardiner:2269}. 
In these expressions, $\hat a$ ($\hat a^\dagger$) and
$\hat b$ ($\hat b^\dagger$)
are the annihilation (creation) operators
for the cavity fields $A$ and $B$, respectively.
We have also defined
$\hat A_{ij} = |i_a\rangle \langle j_a|$ ($i,j = 0, 1$),
and $\hat B_{ij} = |i_b\rangle \langle j_b|$ ($i,j = 0, 1$).
In addition, $g_k$ is the atom-cavity coupling
constant and $\kappa_k$ the cavity bandwidth,
and the phase $\phi$ is related 
to the spatial separation between the source and the target,
cf.~\cite{Carmichael2}.
The jump operators $\hat J_i$ are defined by
\begin{equation}
\hat J_1 = \sqrt{\kappa_a} \hat a + \sqrt{\kappa_b} e^{-i\phi}
\hat b \, ,
\label{jump_1}
\end{equation}
which describes photon emission by the cavities;
\begin{equation}
\hat J_2 = \sqrt{\kappa_a'} \hat a \, , ~~~~~~~
\hat J_3 = \sqrt{\kappa_b'} \hat b \, ,
\label{jumps_2_3}
\end{equation}
are associated with photon absorption or scattering
by the cavity mirrors; and
\begin{equation}
\hat J_4 = \sqrt{\Gamma_a} \hat A_{01} \, , ~~~~~~~
\hat J_5 = \sqrt{\Gamma_b} \hat B_{01} \, ,
\label{jumps_4_5}
\end{equation}
are related to spontaneous emission by the atoms.
Here $\kappa_k'$ and $\Gamma_k$
are the cavity mirrors' absorption (or scattering) rate
and the spontaneous 
emission rate of the two-level atom, respectively.
Note that the operator $\hat J_1$ contains 
the superposition of the two fields radiated by the
two cavities, due to the fact that the radiated photon
cannot be associated with photon emission from 
either $A$ or $B$ separately.

To evaluate the time evolution of the system 
we use a quantum trajectory
approach~\cite{Carmichael2,Dalibard:580,Dum:4382}.
Let us consider the system prepared at time
$t_0 = 0$ in the state $|a\rangle \equiv |1,0,0,0\rangle$,
which denotes the atom $A$ in the state $|1_a\rangle$, 
the cavity $A$ in the vacuum state, the atom $B$ in the
state $|0_b\rangle$ , and the cavity $B$ in the vacuum state.
Similarly, we define $|b\rangle \equiv |0,1,0,0\rangle$,
$|c\rangle \equiv |0,0,1,0\rangle$,
$|d\rangle \equiv |0,0,0,1\rangle$, and
$|e\rangle \equiv |0,0,0,0\rangle$.
To determine 
the state vector of the system at a later time $t$, 
assuming that no jump has occurred between time $t_0$ and $t$,  
we have to solve the nonunitary Schr\"{o}dinger
equation
\begin{equation}
i \hbar \frac{d}{dt} 
| \bar{\psi}_{\rm no} (t) \rangle  =\hat{H^{'}} \, | \bar{\psi}_{\rm no} (t) \rangle \, ,
\label{eq:schr_non_unitary} 
\end{equation}
where $\hat{H^{'}}$ is the non-Hermitian Hamiltonian
given by
\begin{eqnarray}
\hat{H^{'}} &=& \hat H
- \frac{i \hbar}{2} \sum_{i=1}^5 \hat J_i^\dagger
\hat J_i = \hat H_A + \hat H_B - i \hbar \Big(
\frac{K_a}{2}
\hat a^\dagger \hat a  \nonumber \\
&+& \frac{K_b}{2} \hat b^\dagger \hat b
+ \frac{\Gamma_a}{2} \hat A_{11} + 
\frac{\Gamma_b}{2} \hat B_{11} + 
\sqrt{\kappa_a \kappa_b} e^{i \phi} \hat b^\dagger \hat a \Big) ,
\label{eq:n_H_Hamiltonian}
\end{eqnarray}
where we have defined
$K_a = \kappa_a + \kappa_a'$ and $K_b = \kappa_b + \kappa_b'$.
If no jump has occurred between time $t_0$ and $t$, the system evolves via Eq.~(\ref{eq:schr_non_unitary}) into the unnormalized state
\begin{equation}
| \bar{\psi}_{\rm no} (t) \rangle = \alpha(t) |a\rangle + \beta(t) |b \rangle + \gamma(t) |c\rangle + \delta(t) |d \rangle \, .
\label{eq:psi_n}
\end{equation}

The evolution governed by the nonunitary Schr\"odinger equation~(\ref{eq:schr_non_unitary})
is randomly interrupted by one of the five 
kinds of jumps $\hat J_i$,
cf. Eqs.~(\ref{jump_1})-(\ref{jumps_4_5}).
If a jump has occurred at time
$t_{\rm J}$, $t_{\rm J} \in (t_0, t]$, the
wave vector is found collapsed into the state $|e \rangle$ due to the action of one of the jump operators
\begin{equation}
\hspace*{-0.5cm} \hat J_{i} \, | \bar{\psi}_{\rm no} (t_{\rm J}) \rangle  \rightarrow |e \rangle   ~~(i = 1, \ldots, 5). \label{eq:jump_op_i} 
\end{equation}
In the problem under study we may have only one jump. Once the system collapses into the state $|e\rangle$, the nonunitary Schr\"odinger equation~(\ref{eq:schr_non_unitary}) 
lets it remain unchanged.
The density operator $\hat \rho (t)$ is then obtained by performing an ensemble
average over the different 
trajectories at time $t$, yielding the statistical mixture
\begin{equation}
\hat \rho (t) =  
| \bar{\psi}_{\rm no} (t) \rangle \langle
\bar{\psi}_{\rm no} (t) |
+ |\epsilon(t)|^2 |e \rangle \langle e| \, ,
\label{eq:rho_t_1}
\end{equation}
where $|\epsilon(t)|^2 \equiv 1 - \langle \bar{\psi}_{\rm no} (t) |\bar{\psi}_{\rm no} (t) \rangle$.
The values 
$|\alpha(t)|^2$, $|\beta(t)|^2$, $|\gamma(t)|^2$, $|\delta(t)|^2$, and $|\epsilon(t)|^2$
represent the probabilities that at time $t$ the system can be found either in $|a\rangle$,
$|b\rangle$, $|c\rangle$, $|d\rangle$, or $|e\rangle$, respectively.

In order to determine $\alpha(t)$, $\beta(t)$,
$\gamma(t)$, and $\delta(t)$, we have to solve
the nonunitary Schr\"odinger equation, cf. Eqs.~(\ref{eq:schr_non_unitary})
and~(\ref{eq:n_H_Hamiltonian}), which
leads to the
inhomogeneous system of differential equations
\begin{equation}
\left\{ 
\begin{array}{llll}
\dot \alpha(t) 
= -i \left( \Delta_a - i \Gamma_a/2 \right)\alpha (t) - i g_a \beta(t) \, , \\
\dot \beta(t) = - i g_a \alpha(t) - (K_a/2) \beta(t) \, , \\
\dot \gamma(t) = -i \left( \Delta_b - i \Gamma_b/2 \right)\gamma (t) - i g_b \delta(t) \, ,  \\
\dot \delta(t) = - i g_b \gamma(t) - (K_b/2) \delta(t) 
-\sqrt{\kappa_a\kappa_b} e^{i \phi} \beta (t) \, .
\end{array}
\right.
\label{eq:sys_diff_eq_1}
\end{equation}
For the initial conditions $\alpha(0) \!=\!1$, 
$\beta(0) \!=\! 0$, $\gamma(0) \!=\!0$, and $\delta(0) \!=\! 0$,
and defining
\begin{equation}
\Omega_k \!\equiv\! \sqrt{\frac{K_k^2}{4} \!-\! 4g_k^2 \!-\! i K_k \!\left(\Delta_k \!-\!i \frac{\Gamma_k}{2}\right) \!-\! \left(\Delta_k \!-\! i \frac{\Gamma_k}{2}\right)^2 } \, , 
\end{equation}
we get, similarly
as done in~\cite{Difidio:032334}, the solutions 
\begin{eqnarray}
{\alpha} (t) &=&  \left[\frac{K_a/2 - i (\Delta_a - i \Gamma_a/2)}{\Omega_a} \sinh \! \left(\frac{\Omega_a t}{2}\right)  \right.
\nonumber \\
&+& \left. \cosh \! \left( \frac{\Omega_a t}{2}\right) \right] \! e^{-[(K_a +\Gamma_a)/4 
+ i \Delta_a/2]t} , \nonumber \\
{\beta} (t) &=& - \frac{2ig_a}{\Omega_a} \sinh \! \left(\frac{\Omega_a t}{2}\right) e^{-[(K_a + \Gamma_a)/4 + i\Delta_a/2]t}\, ,\nonumber \\
{\gamma} (t) &=& g_b \left\{
f_{+}(t) \! \left[g_{-}(t) \!+\! h_{+}(t)  \right] 
\!-\! f_{-}(t) \!
\left[g_{+}(t) \!+\! h_{-}(t)  \right] \right\} ,
\nonumber \\
{\delta} (t) &=& 
i \! \left[\frac{K_b \!-\! \Gamma_b}{4} \!-\! 
i \frac{\Delta_b}{2} \!+\! \frac{\Omega_b}{2}\right] \! f_{-}(t) 
\! \left[g_{+}(t) \!+\! h_{-}(t)  \right] \nonumber \\
&-& i \! \left[\frac{K_b \!-\! \Gamma_b}{4} \!-\!
i \frac{\Delta_b}{2} \!-\! \frac{\Omega_b}{2}\right] \! f_{+}(t) 
\! \left[g_{-}(t) \!+\! h_{+}(t)  \right] \!.
\label{eq:diff_eq_sol_general_g_d}
\end{eqnarray}
Here we have defined,
\begin{equation}
f_\pm(t) = \frac{g_a \sqrt{\kappa_a\kappa_b} e^{i \phi} }{\Omega_a \Omega_b} e^{[-(K_b + \Gamma_b)/4 - i \Delta_b/2 \pm \Omega_b/2]t} \, ,
\label{eq:fpm}
\end{equation}
\begin{equation}
g_\pm(t) = \frac{e^{[(\Omega_a\pm \Omega_b)/2 - \Upsilon - i \Lambda]t}-1}{(\Omega_a\pm \Omega_b)/2 - \Upsilon - i \Lambda} \, ,
\end{equation}
and
\begin{equation}
h_\pm(t) = \frac{e^{-[(\Omega_a\pm \Omega_b)/2 + \Upsilon + i \Lambda]t}-1}{(\Omega_a\pm \Omega_b)/2 + \Upsilon + i \Lambda} \, ,
\end{equation}
where $\Upsilon \!=\! (K_a \!-\! K_b \!+\! \Gamma_a \!-\! \Gamma_b)/4$ and $\Lambda \!=\! (\Delta_a \!-\! \Delta_b)/2$.
In the case of equal parameters for the
two subsystems $A$ and $B$, the solutions
for $\gamma(t)$ and $\delta(t)$
simplify to
\begin{eqnarray}
\gamma (t) \!\!&=&\!\! \frac{2\kappa g^2 e^{i \phi}}{\Omega^3}
\Big[ \Omega t \cosh \! \left(\!\frac{\Omega t}{2}\!\right) 
- 2 \sinh \! \left(\!\frac{\Omega t}{2}\!\right) \Big]\nonumber \\
\!\!&\times&\!\! e^{-[(K + \Gamma)/4 + i \Delta/2]t} \, , 
\nonumber \\
\delta (t) \!\!&=&\!\! \frac{i\kappa g e^{i\phi}}{\Omega^3}
\Big\{ \! \Big( \frac{K \!-\! \Gamma}{2} \!-\! i \Delta \Big) \!
\Big[ 2 \sinh \! \left(\!\frac{\Omega t}{2}\!\right) \!-\! \Omega t 
 \cosh \! \left(\!\frac{\Omega t}{2}\!\right) \! \Big] \nonumber \\  
\!\!&+&\!\! \Omega^2 t \sinh \! \left(\!\frac{\Omega t}{2}\!\right) \!
\Big\}\, e^{-[(K + \Gamma)/4 + i \Delta/2]t}\, ,
\label{eq:diff_eq_sol_equal_values}
\end{eqnarray}
where
$\kappa \!=\! \kappa_a \!=\! \kappa_b$,
$K\!=\! K_a \!=\! K_b$,
$\Delta \!=\! \Delta_a \!=\! \Delta_b$, 
$\Gamma \!=\! \Gamma_a \!=\! \Gamma_b$, $g \!=\! g_a \!=\! g_b$,
and $\Omega \!=\! \Omega_a \!=\! \Omega_b$.

In the system under study, because only one
atom is initially excited,
the two intracavity fields constitute a pair of
entangled qubits, for a detailed discussion
of single-particle entanglement, see~\cite{Enk:064306}.
An appropriate measure of the entanglement
for a two-qubit system is the concurrence~\cite{Wootters:2245}.
To derive an expression for the  
concurrence between the two intracavity fields
we consider the density operator obtained 
by tracing over the atomic states for 
the two subsystems, $\hat \rho_{\rm cav}(t) \!=\! {\rm Tr}_{\rm at} 
\left[ \hat \rho (t) \right]$.
It is easy to show,
following Ref.~\cite{Wootters:2245}, that the concurrence
between the two intracavity fields is given by
\begin{equation}
C[\rho_{\rm cav}(t)] = 2 \left| \beta(t) \right| \left| 
\delta(t)  \right| \, .
\label{eq:conc_cavities}
\end{equation}
Note that for equal parameters for the two subsystems, and
for $g \gg K, \Gamma, \Delta$, the concurrence is
given by $C[\rho_{\rm cav}(t)]\simeq \kappa t \sin^2(gt) e^{[-(K+\Gamma)t/2]}$.

Following~\cite{Blow:4102,Legero:797}, we consider 
a photon in the mode $\xi_i$,
the mode escaping from the cavities and going to the photodiode PD.
It is described by the normalized function $\xi_i(t)$ 
of amplitude
envelope $\zeta_i(t)$ and phase $\phi_i(t)$,
$\xi_i (t) = \zeta_i(t) e^{i \phi_i(t)}$,
with 
\begin{equation}
\int_{0}^{\infty} dt \, |\xi_i (t)|^2 =
\int_{0}^{\infty} dt \, \zeta^2_i(t) = 1 \, .
\label{eq:norm_xi}
\end{equation} 
When a photon is in the
mode $\xi_i$, whose amplitude envelope $\zeta_i(t)$
does not change significantly in
the detection time resolution $T$,
the response probability of the detector of quantum efficiency $\eta$
within a time interval $[t-T/2, t+T/2]$ is given by~\cite{Difidio:043822}
\begin{equation}
P_{\rm D}(t)= \eta \, p_{\rm rad}(\infty) 
\zeta^2_i(t) \, T \, .
\label{eq:P_i_D_eta}
\end{equation}
Here $p_{\rm rad}(\infty)=\lim_{t \to \infty} p_{\rm rad}(t)$,
where the function $p_{\rm rad}(t)$ represents 
the probability that a photon
is radiated by the cascaded system 
in the time interval $[0, t]$, 
which reads as
\begin{eqnarray}
p_{\rm rad}(t) \!\!&=&\!\! \!\! \int_0^t \! dt' 
\langle \hat J_1^\dagger \hat J_1  \rangle_{t'}
\!=\! \kappa_a \!\! \int_0^t \! dt'  |\beta(t')|^2 
\!+\! \kappa_b \!\! \int_0^t \!dt'  |\delta (t')|^2 \nonumber \\
\!&+&\! 2 \, \sqrt{\kappa_a \kappa_b} \int_0^t \!
dt' {\rm Re} \! \left[ \beta^{*}(t') \delta (t')e^{-i\phi} \right]\, .
\label{eq:p_rad_t}
\end{eqnarray}
Since $\delta (t)$ contains an
overall factor $e^{i\phi}$, cf. Eqs.~(\ref{eq:diff_eq_sol_general_g_d}) and (\ref{eq:fpm}), the phase $\phi$ is irrelevant in Eq.~(\ref{eq:p_rad_t}).

\begin{figure}
\includegraphics[width=7.5cm]{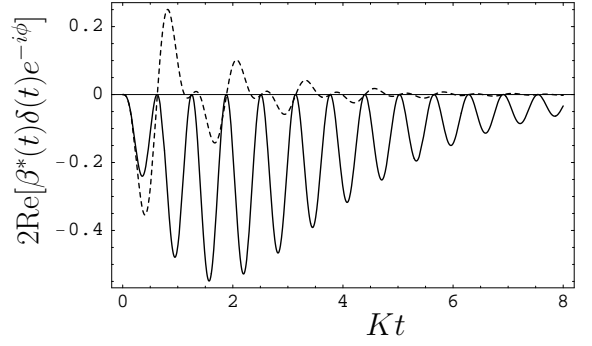}
\caption{The function $2 {\rm Re}[\beta^{*}(t)\delta(t)e^{-i\phi}]$ 
is shown for equal parameters for the
two subsystems $A$ and $B$, for $g/K = 5$,
$\kappa/K = 0.9$, $\Delta/K = 0.1$,
$\Gamma/K = 0.2$ (solid line), 
when Eq.~(\ref{eq:concurrence-approx}) applies.
The case 
when no atom is present in the second cavity, i.e. $g_b=0$, 
is also shown (dashed line).}
\label{fig:figure_2}
\end{figure}
The probability to measure between time
$t - T/2$ and $t + T/2$ a ``click" at the detector 
is equal to the probability to have a jump $\hat J_1$ in the
same time interval, so that using 
Eq.~(\ref{jump_1}), we get
\begin{eqnarray}
P_{\rm D} (t) &=& \eta {\rm Tr} \left[ \hat \rho (t)  \hat J_1^\dagger \hat J_1 \right] T = 
\eta T \Big\{ \kappa_a |\beta(t)|^2 + \kappa_b |\delta(t)|^2 \nonumber \\
&+& 2 \sqrt{\kappa_a \kappa_a} \, {\rm Re}[\beta^{*}(t)\delta(t)e^{-i\phi}] \Big\}\, .
\label{eq:D_jumps}
\end{eqnarray}
Comparing this with
Eq.~(\ref{eq:P_i_D_eta}) we obtain
\begin{equation}
\zeta^2_i(t) = \frac{\kappa_a |\beta(t)|^2 + \kappa_b |\delta(t)|^2
+ 2 \sqrt{\kappa_a \kappa_b} \, {\rm Re}[\beta^{*}(t)\delta(t)e^{-i\phi}] }{p_{\rm rad}(\infty)}\, .
\label{eq:epsilon_i_1}
\end{equation}
Note that Eq.~(\ref{eq:norm_xi}) is correctly fulfilled.

Let us now analyze in more details the term 
$2  {\rm Re}[\beta^{*}(t)\delta(t)e^{-i\phi}]$
in Eq.~(\ref{eq:epsilon_i_1}).
Writing $\beta(t) = |\beta(t)|e^{i \phi_{\beta}(t)}$
and $\delta(t) = e^{i\phi}|\delta(t)|e^{i \phi_{\delta}(t)}$,
yields
\begin{equation}
2 {\rm Re}[\beta^{*}(t)\delta(t)e^{-i\phi}]
\!=\! C[\rho_{\rm cav}(t)] \cos\left[ \phi_{\delta}(t) - \phi_{\beta}(t) \right]
\,,
\label{eq:interference}
\end{equation}
where $C[\rho_{\rm cav}(t)]$
is the concurrence between the two intracavity fields,
cf. Eq.~(\ref{eq:conc_cavities}).
In this respect, Eq.~(\ref{eq:epsilon_i_1})
clearly shows that the mode structure of the radiated
field depends not only on the two
intracavity fields, i.e. $|\beta(t)|^2$ and $|\delta(t)|^2$,
but also on the entanglement established between them.
This represents an interference between the 
possibility to have the photon in one or in the other cavity.
For equal parameters for the two subsystems, and
for $g \gg K, \Gamma, \Delta$, one obtains the relation
\begin{equation}
2 {\rm Re}[\beta^{*}(t)\delta(t)e^{-i\phi}]\simeq - C[\rho_{\rm cav}(t)] \, .
\label{eq:concurrence-approx}
\end{equation}
%Note that 
In this case the concurrence can be
experimentally 
derived by using the combination of two measurements. The first
one, by using only cavity $A$, gives $|\beta(t)|$, via the relation
$P_{\rm D}'(t) = \eta \kappa T |\beta(t)|^2$, cf.~\cite{Difidio:043822}.
The second measurement, by using both cavities, gives
$|\delta(t)|$ via the relation
$P_{\rm D}(t)/P_{\rm D}'(t)=(1 - |\delta(t)|/|\beta(t)|)^2$. Knowing $|\beta(t)|$ and $|\delta(t)|$,
the concurrence is obtained from Eq.~(\ref{eq:conc_cavities}).

In Fig.~\ref{fig:figure_2} we show the term
$2 {\rm Re}[\beta^{*}(t)\delta(t)e^{-i\phi}]$
under conditions when it represents the
negative concurrence according to Eq.~(\ref{eq:concurrence-approx}).
We also show the case
when no atom is present in the second cavity. In both cases
the entanglement between the
two intracavity fields gives a significant contribution to
the mode structure of the radiated photon, cf. Eq.~(\ref{eq:epsilon_i_1}).

\begin{figure}
\includegraphics[width=7.5cm]{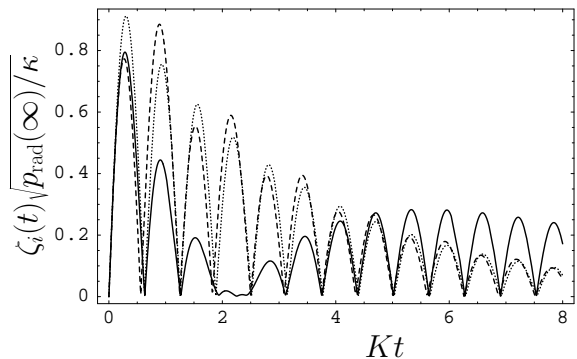}
\caption{The amplitude envelope
$\zeta_i(t)\sqrt{p_{\rm rad}(\infty)/\kappa}$
for the mode of the radiated field is shown
for equal parameters for the
two subsystems $A$ and $B$, for $g/K = 5$,
$\kappa/K = 0.9$, $\Delta/K = 0.1$,
$\Gamma/K = 0.2$ (full line),
and for no atom in the second cavity (dashed line).
The case where the subsystem $B$ is absent,
i.e. for $K_b=0$, is also shown (dotted line).}
\label{fig:figure_3}
\end{figure}
The amplitude envelope for the mode of the radiated field
is shown in Fig.~\ref{fig:figure_3}, 
for equal parameters of the
two subsystems and for the case
when no atom is present in the second cavity.
The case when the subsystem $B$ is absent,
i.e. for $K_b=0$, is also shown, reproducing
the result obtained in~\cite{Difidio:043822}.
The shown mode structures carring
the entanglement signature
could be realized and observed by
extending the experimental setup described in~\cite{Bochmann:223601}.
By measuring the arrival time distribution of the photon radiated from a
system with equal cavity parameters, one may determine the full dynamics of
the concurrence and hence the entanglement dynamics of the two 
intracavity fields in the strong coupling regime. 
This regime has been realized in recent experiments~\cite{Wilk:488,Kimble:1447}.

In conclusion, the dynamics of a system consisting
of two atom-cavity subsystems has been
analyzed under realistic conditions with losses. 
%The mode structure of a single photon has been analyzed.
For properly chosen parameters, the mode
function of the single photon escaping from the cavities
reflects the full dynamics of the concurrence of the two intracavity fields,
while they continue to interact with the two atoms.
This allows one to
%may be useful for 
detect the entanglement dynamics of two
cavity fields, and may be useful for transferring the information on entanglement by a single photon over a large distance.

This work was supported by the Deutsche Forschungsgemeinschaft.

\end{document}